\documentclass[journal]{IEEEtran}
\usepackage{amsmath,amsfonts}
\usepackage{multirow}
\usepackage{algorithm}
\usepackage{algorithmic}
\usepackage{graphicx}
\usepackage{textcomp}
\usepackage{xcolor}
\usepackage{framed}
\usepackage{listings}
\usepackage{bbding} 
\usepackage{color}
\usepackage{enumitem}
\usepackage{threeparttable}
\usepackage{multicol}
\usepackage{tabularx}
\usepackage{enumitem}
\usepackage{booktabs}
\usepackage{xspace}
\usepackage{hyperref}
\usepackage{cleveref}
\usepackage[most]{tcolorbox} 

\makeatletter
\def\UrlAlphabet{%
      \do\a\do\b\do\c\do\d\do\e\do\f\do\g\do\h\do\i\do\j
      \do\k\do\l\do\m\do\n\do\o\do\p\do\q\do\r\do\s\do\t
      \do\u\do\v\do\w\do\x\do\y\do\z\do\A\do\B\do\C\do\D
      \do\E\do\F\do\G\do\H\do\I\do\J\do\K\do\L\do\M\do\N
      \do\O\do\P\do\Q\do\R\do\S\do\T\do\U\do\V\do\W\do\X
      \do\Y\do\Z\do\*\do\-\do\'\do\"\do\_\do\.}
\def\UrlDigits{\do\1\do\2\do\3\do\4\do\5\do\6\do\7\do\8\do\9\do\0}
\g@addto@macro{\UrlBreaks}{\UrlOrds}
\g@addto@macro{\UrlBreaks}{\UrlAlphabet}
\g@addto@macro{\UrlBreaks}{\UrlDigits}
\makeatother

\tcbuselibrary{breakable}
\def\BibTeX{{\rm B\kern-.05em{\sc i\kern-.025em b}\kern-.08em
    T\kern-.1667em\lower.7ex\hbox{E}\kern-.125emX}}
\usepackage{balance}

\begin{document}

\title{Effective Construction of Feature Database of Fingerprinting Functions for Code-Clone SCA}

\author{Zhiling Zhu, Tieming Chen, Chengwei Liu, Jiahui Wu, Qijie Song, \\ Zhengzi Xu, Yueming Wu, and Yang Liu
\thanks{Chengwei Liu and Tieming Chen are corresponding authors of this work.}
\thanks{Zhiling Zhu, Tieming Chen and Qijie Song are with Zhejiang University of Technology, China. (e-mail: zhilingzhu@zjut.edu.cn, tmchen@zjut.edu.cn, songqijie@zjut.edu.cn).}
\thanks{Chengwei Liu, Jiahui Wu and Yang Liu are with Nanyang Technological University, Singapore (e-mail: chengwei.liu@ntu.edu.sg, jiahui004@e.ntu.edu.sg, yangliu@ntu.edu.sg).}
\thanks{Zhengzi Xu is with Imperial Global Singapore and Imperial College London (e-mail:z.xu@imperial.ac.uk)}
\thanks{Yueming Wu is with Huazhong University of Science and Technology, China (e-mail: wuyueming21@gmail.com).}}

\markboth{Preprint}{}

\maketitle

\begin{abstract}
In contemporary software development, the extensive adoption of third-party libraries has become a standard practice to avoid reinventing the wheel. This trend is particularly evident in programming ecosystems that lack mature package managers, such as C/C++. Within this context, Code Clone Detection (CCD) techniques have emerged as essential tools for identifying instances of code reuse, thereby mitigating security and legal concerns. Despite significant scholarly advancements aimed at enhancing the accuracy of CCD algorithms, a notable gap persists between the capabilities of academic tools and the demands of real-world applications. This disparity primarily arises from the reliance on academic tools on experimental datasets composed of a limited number of indexed repositories, whereas real-world applications require a much broader scope of code sources.

To address this challenge, we propose a novel, systematic approach for selecting appropriate open-source repositories to construct a more effective CCD feature dataset. Our method, OSSScope, employs a greedy algorithm to analyze the diversity of fingerprinting functions across 2.55 million C repositories on GitHub. The objective is to maximize the inclusion of unique fingerprinting functions while minimizing the number of repositories required. OSSScope successfully compiles an optimal feature dataset containing 61 million fingerprinting functions from 190K repositories, resulting in a 22.65\% improvement in recall over SOTA tools. Moreover, the constructed dataset covers over 95\% of the features from other sources, demonstrating its diversity and coverage.
Furthermore, leveraging the repositories selected by OSSScope, we introduce RClassifier, a lightweight classification model that streamlines the construction process for migration to other language ecosystems. By capturing key metrics, RClassifier directly determines whether a repository should be included in the dataset. Our comprehensive evaluation of RClassifier within the Golang ecosystem validates its effectiveness, achieving recall performance comparable to the feature dataset produced by OSSScope.

\end{abstract}


\section{Introduction}
\label{sec:Introduction}
\IEEEPARstart{A}{s} the extensive reuse of open-source artifacts in modern software development prevents reinventing wheels, software composition analysis (SCA) techniques have been widely populated to detect the reuse of third-party source code to avoid potential risks, such as security vulnerabilities and license compliance, etc. 
Apart from the reuse of third-party libraries as dependencies, which could be easily identified by corresponding package managers (e.g., \texttt{Maven} for Java, \texttt{NPM} for JavaScript), direct copy-paste of code fragments from existing software repositories is a much more common practice in software ecosystems that have no popular package managers, such as C/C++ ecosystem~\cite{tang2022towards}. To this end, code clone detection (CCD) techniques have been introduced to detect such copy-paste reuse of third-party code fragments.

CCD tool usually consists of a pre-constructed feature dataset and a well-designed code clone detection algorithm. Typically, CCD tools report potential code clones by comparing target projects with feature databases based on the code clone detection algorithms. 
Many CCD tools have been proposed to improve their accuracy and performance by strengthening code clone detection algorithms. 
For instance, 
SourcererCC~\cite{sajnani2016sourcerercc} and Vuddy~\cite{vuddy} employed code tokens as the representations of features to enable scalable code clone detections, Yang et al.~\cite{yang2018structural} introduced the AST representation and adopted Smith-Waterman algorithm to enhance the precision, CCSharp~\cite{wang2017ccsharp} further improved code clone detection by utilizing program dependency graph for semantic level similarity, CENTRIS~\cite{woo2021centris} adopted locality-sensitive hashing to mapping type 1 and type 2 code clones for recognizing the integrated TPLs, and OSSFP~\cite{wu2023ossfp} further improved the accuracy by filtering the fingerprinting functions for similarity calculation. 
However, these tools are only implemented and evaluated with feature datasets derived from limited open-source code to demonstrate the capability, e.g., roughly selecting C/C++ repositories with over 100 stars in SOTA tools (CENTRIS~\cite{woo2021centris} and OSSFP~\cite{wu2023ossfp}).
while overlooking that limited datasets could also heavily impact the generalization of SCA tools in real-world applications, as revealed by Jiang et al.~\cite{tplite}. 
More specifically, since it is impossible to detect if cloned codes are not included in the feature dataset, these existing works usually evaluate algorithms and tools by ground truths that are included in the scope of their experimental feature datasets, which overlook the case that code clones could happen on a much larger range of open-source repositories, leading to the compromised applicability in real-world SCA detection.
Moreover, commercial tools choose to collect all available source code (i.e., BlackDuck~\cite{bdknowledge} had collected 6.3 million components as its feature database~\cite{bd-feature}) to ensure the recall, but this, on the other hand, could be resource-consuming and result in a bloated feature dataset that could compromise the performance of CCD tools~\cite{wu2022detecting,wu2020scdetector}, which makes it difficult to employ complex CCD algorithms in these commercial tools.
Therefore, it is non-trivial to construct effective-efficient datasets for code clone-based SCA tools. We aim to explore the proper methodology to construct such effective-efficient datasets for code clone detection.

To achieve this, we need to balance the trade-offs between the coverage and high quality of feature datasets. 
\textbf{1) Code Coverage.} There are tens of millions of open source repositories on open source repositories; it would be resource-consuming to construct feature datasets for all source codes of these repositories. Moreover, considering that repositories could be updated constantly, the various versions of historical code could exponentially increase the difficulties when constructing feature datasets. 
\textbf{2) Code Clone Efficiency.} There could be a significantly large portion of source codes that are not necessarily included in the feature dataset (not IP-significant~\cite{wu2023ossfp}), these unnecessary code features should be further excluded to ensure the efficiency of code clone detection. However, there is a lack of sufficient and comprehensive solutions for such feature dataset refinement.

To balance the trade-off, we first take the ecosystem of C language as an example to conduct a comprehensive investigation on the diversity of C code, as well as the preference of code clones in the C ecosystem. Specifically, we first propose a novel approach to explore the proper scope of C repositories from the perspective of including all identical C functions. Since it is a huge consumption of time and resources to directly download all existing C repositories and parse the code in all versions, we propose a novel approach to greedily construct the feature dataset by including as many unique functions as possible while retrieving minimal numbers of repositories. After reaching the theoretic upper bound number of unique functions, we exclude repositories that contribute little to code clone detection by three proposed metrics on repository values, to approximate the optimal feature dataset (i.e., OSSScope) of CCD with balanced precision and performance. 
As a result, after analyzing 385,000 repositories, OSSScope identified and included 73,135,396 unique functions, accounting for 95\% of the theoretical upper bound of unique functions as predicted by curve fitting. Furthermore, after filtering out repositories with low contributions to CCD, we obtained an effective and efficient feature dataset comprising 61,746,893 unique functions (79.52\% of all identified functions) from 193,026 repositories, which represent only 7.57\% of all C repositories on GitHub. Our evaluation demonstrated that OSSScope significantly enhances the recall of existing CCD tools by 22.65\% while incurring only a minimal performance overhead. This result highlights the superior balance achieved by OSSScope between accuracy and computational efficiency compared to existing feature dataset construction strategies. Additionally, the constructed dataset covers over 95\% of the features from other sources, further validating its diversity and comprehensive representation.

Furthermore, although OSSScope is effective for feature dataset construction, it is still resource-consuming to analyze millions of repositories when generalized to other language ecosystems. We turned the repository selection into a binary classification problem, and trained a classification model, named RClassifier, based on the light-weighted and easy-obtained metrics. Our validation on the Golang ecosystem showed that the feature dataset constructed from repositories sorted out by RClassifier can reach 86.69\% precision and 89.67\% recall, which is only slightly lower than the recall (92.54\%) of the dataset constructed by OSSScope.

In summary, the main contributions are as follows:
\begin{itemize}[leftmargin=*]
    \item We proposed OSSScope, a resource-consuming but very effective approach to construct a code clone feature database from the open-source repositories. 
    \item Using OSSScope, we built an effective CCD feature dataset for C, containing 62M unique functions from 190K repositories (80.2\% of unique functions from just 7.57\% of GitHub C repositories), achieving a 22.65\% recall improvement over SOTA CCD tools. Covering over 95\% of features from other sources, the dataset demonstrates strong diversity and comprehensive coverage.
    \item We proposed RClassifier to construct CCD feature datasets using easily obtainable repository metrics. Evaluations on the Golang ecosystem showed that RClassifier effectively selects repositories with only a slight recall reduction (3\%) compared to OSSScope, demonstrating its generalizability across language ecosystems.
    \item We have open-sourced the constructed feature dataset of C and Golang on our website~\cite{public-data}. To our best knowledge, so far, they are the largest open-source feature databases for code clone detection for C and Golang.
\end{itemize}

\section{Related Works}
\label{sec:background}

\subsection{Software Composition Analysis and Code Clone Detection}

Software Composition Analysis (SCA) tools play a crucial role in identifying third-party libraries (TPLs) in software projects to enhance security and compliance. Two primary approaches to TPL detection exist: SBOM (software bill of materials)~\cite{sbom} comparisons and Code Clone Detection (CCD). In the C/C++ ecosystem, where the absence of a standardized SBOM renders SBOM-based detection ineffective, CCD-based methods are predominantly used. Several CCD-based SCA tools have been developed to improve TPL detection across different ecosystems. For instance, LibScout~\cite{backes2016reliable} identifies third-party libraries in Android collected from the Google Play Store and third-party markets. Similarly, LibPecker~\cite{zhang2018detecting} detects libraries in Android apps using signature matching techniques. OSSFP~\cite{wu2023ossfp} advances TPL detection for C/C++ projects by applying function fingerprinting. Other tools, such as CENTRIS~\cite{woo2021centris} and V1SCAN~\cite{v1scan}, focus on identifying vulnerabilities in reused open-source software (OSS) by analyzing code reuse in popular repositories. These tools utilize various datasets for validation. LibScout~\cite{backes2016reliable} manually collects data from 164 popular Android applications, while LibPecker~\cite{zhang2018detecting} relies on datasets curated from top Android libraries~\cite{top100,awesome}. OSSFP~\cite{wu2023ossfp}, CENTRIS~\cite{woo2021centris}, and V1SCAN~\cite{v1scan} use repositories with at least 100 GitHub stars, resulting in a feature dataset of approximately 12K repositories. This selective approach ensures that only widely used libraries are analyzed, optimizing detection accuracy and minimizing false positives.

However, the limited scope and simplistic selection criteria of existing evaluation datasets pose a threat 
that such evaluation can not evaluate the real-world application scenarios in which there are far more open-source projects for reference and reuse.
By focusing solely on limited scopes of libraries or repositories, these tools risk overlooking a substantial number of libraries that share similar code signatures but fall outside the selected dataset. This limitation can result in false negatives, as less popular or niche libraries may remain undetected, ultimately compromising the accuracy and real-world applicability. Consequently, there remains a lack of a systematic approach to enhancing the applicability of CCD-based SCA tools from the perspective of feature dataset construction.

\subsection{Survey on Metrics Selection Criteria for Open Source Repository}

To this end, we further conducted a small literature review on existing open-source related papers and investigated from which aspects they select their representative datasets.
Given that selecting open-source repositories for feature dataset construction falls within the scope of Mining Software Repositories (MSR), we selected \textit{\{"Open Source Software" OR "OSS"\} AND \{"Mining Software Repository" OR "MSR" \}} in abstract and within five years as search criteria, following the guidance as the previous SLR studies ~\cite{nightingale2009guide,kitchenham2009systematic}. We selected DBLP~\cite{dblp}, ACM Digital Library ~\cite{acm}, IEEE Xplore ~\cite{ieee}, and Elsevier ScienceDirect ~\cite{elsevier} as sources and obtained 80 papers as the starting point. After mannully examed the content, we sorted out 15 papers that mentioned metrics for repository filtering. Then, we snowballed their referencens to find other potential related papers ~\cite{wohlin2014guidelines}. After two rounds snowballing, we further identified 14 papers. The final set of the SLR contains 29 papers.

\begin{table}[t]
    \caption{Metrics Used in Mining Open Source Software}
    \label{tab:metrics-summary}
    \footnotesize
    \begin{tabularx}{\linewidth}{cX}
        \toprule
            Metrics     & Mentioned Papers \\ \midrule
            stars       &  \cite{yang2023users,wang2022vcmatch,yan2023github,hasabnis2022gitrank,santos2023tell,hasan2023understanding,foalem2024studying,mastropaolo2023robustness,bg-3,bg-4,bg-5,bg-6,bg-7,bg-8,bg-10,bg-11,bg-12,bg-13,bg-14,albonico2021mining}                \\
            forks       &   \cite{yang2023users,yan2023github,hasabnis2022gitrank,wang2023fork,bg-3,bg-6,bg-9}        \\
            watcher     &   \cite{yang2023users,hasabnis2022gitrank,albonico2021mining}  \\
            issues       &   \cite{yang2023users,yan2023github,hasabnis2022gitrank,santos2021can,ehsani2024incivility,vargovich2023givemelabeledissues,heo2023empirical,wang2023fork,bg-3,bg-4,bg-6,bg-11,bg-12,bg-14}        \\
            commits       &  \cite{yan2023github,foalem2024studying,calefato2024lot,deng2021changes,mastropaolo2023robustness,bg-6,bg-11,albonico2021mining}         \\
            update time  &  \cite{yang2023users,ehsani2024incivility,santos2023tell,deng2021changes,bg-9,bg-15}    \\
            contributors  &  \cite{ehsani2024incivility,foalem2024studying,calefato2024lot,deng2021changes,mastropaolo2023robustness,bg-10,bg-11,albonico2021mining}      \\
            PR  &        \cite{vargovich2023givemelabeledissues,santos2023tell,hasan2023understanding,calefato2024lot,wang2023fork,bg-6,bg-11,bg-14}          \\
        \bottomrule
    \end{tabularx}
\end{table}

All utilized metrics are summarized in \Cref{tab:metrics-summary}. The adopted metrics can be classified into three primary dimensions: project popularity (e.g., number of stars, forks, etc.), project activity (e.g., number of contributors, active duration, etc.), and user concerns (e.g., number of issues, pull requests, etc.). These metrics collectively characterize repositories from multiple perspectives. Repositories that score well across these metrics are more likely to gain user attention and, consequently, have a higher probability of being reused. Therefore, 
we take these metrics as the primary indicators for our greedy strategy of SCA feature dataset construction in~\Cref{sec:exploratory}.

\section{Exploratory Study on C Ecosystem}
\label{sec:exploratory}

\begin{figure*}[t]
    \centering
    \includegraphics[width=0.95\textwidth]{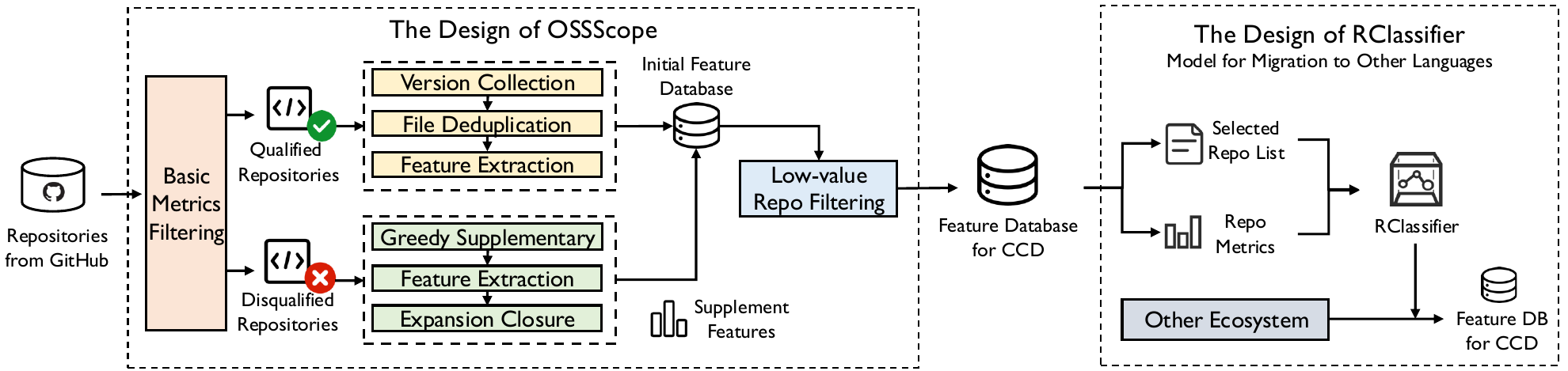}
    \caption{Overviews of the proposed methods}
    \label{fig:overviews}
\end{figure*}

In this section, we first introduce the proposed framework \textbf{OSSScope} to construct an effective-efficient feature dataset on C repositories from GitHub as an exploratory study. Generally, the basic idea is to find out and include repositories that are most likely to be copied by others, and it includes four major steps: 1) construct the fundamental feature dataset by qualified repositories, 2) expand the feature dataset by including disqualified repositories greedily in batches, 3) expansion closure, and 4) filtering out less contributed repositories. Based on the selected repositories with metrics, we trained a lightweight model, \textbf{RClassifier}, for migrating to other ecosystems, which are discussed at ~\Cref{sec:migration}. The overview of the proposed methods is presented in \Cref{fig:overviews}. 

\subsection{Metric-based Filtering of Repositories}

Given the substantial computational resources required to directly compute and analyze features for each of the more than 2.5 million C repositories (as of June 1, 2023), we adopted a strategic, resource-efficient approach. Specifically, we implemented a pre-selection process based on established practices outlined in ~\Cref{sec:background}. This process aimed to identify repositories with a higher likelihood of being duplicated or widely used by other users before conducting intensive feature computation and constructing the feature dataset.

To ensure the efficiency and feasibility of this pre-selection process, we prioritized metrics that are readily accessible from GitHub metadata. Consequently, we identified six metrics, categorized into three distinct aspects, to serve as proxies for estimating the likelihood of a repository being reused. Additionally, the repository size must be greater than zero to prevent empty items.

\begin{itemize}[leftmargin=*]
    \item \textbf{Popularity (Stars \& Forks)}: We incorporate the number of stars and forks as key indicators of a repository's popularity. Repositories that attract more stars and forks are more likely to be sources of code that are copied and reused, reflecting a broader recognition and validation by the community. 
    \item \textbf{User Concerns (Issues)}: The volume of associated issues serves as a proxy for community interest and concern. A higher number of open issues suggests a more engaged user base, potentially indicating areas where code reuse is more probable.
    \item \textbf{Liveness (Commits, Update Times \& Contributors)}: Indicators of activity and maintenance collectively suggest a repository's longevity. Repositories demonstrating sustained activity over time are presumed to have a greater influence and are more attractive for developers to reuse their code.
\end{itemize}

\begin{table}[t]
    \centering
    \caption{Criteria of the Metrics-based Filtering}
    \label{tab:criteria}
    \footnotesize
    \scalebox{0.8}{
    \begin{tabular}{l|rl|l|rl}
        \toprule
        Characteristic              & Metrics      & Criteria        & Characteristic            & Metrics       & Criteria        \\
        \midrule
        \multirow{2}{*}{Popularity} & \#Star       & \textgreater{}0 & \multirow{4}{*}{Liveness} & \#Commit      & \textgreater{}1 \\
                                    & \#Fork       & \textgreater{}0 &                           & Action Period & N.A.            \\
        \cmidrule{0-2}
        User Concerns               & \#Open Issue & \textgreater{}0 &                           & \#Contributor   & \textgreater{}1 \\
        \cmidrule{0-2}
        Size                        & Repo Size    & \textgreater{}0 &                           &               &                 \\
        \bottomrule
    \end{tabular}}
\end{table}

To rigorously ensure the inclusion of repositories with a high likelihood of being copied, we deliberately established minimal filtering criteria, as detailed in ~\Cref{tab:criteria}. Specifically, for a repository to qualify for our initial feature dataset, it must have at least one star, one fork, one open issue, and at least one commit from contributors other than the repository owner. Additionally, we analyze the temporal dynamics of repository engagement by measuring the duration from a repository's creation to its most recent activity, referred to as the action period. These thresholds are intentionally set low to retain as many repositories as possible that might be copied by other users. Repositories that do not meet these criteria, indicating a lack of user engagement or a short lifespan, are less likely to be widely copied. Nevertheless, to mitigate the risk of excluding potentially relevant repositories, we have implemented a supplementary step as follows to reintegrate items that may have been inadvertently filtered out by these initial criteria.

Following the application of these filtering criteria, a total of 62,868 repositories were identified as constituting the preliminary dataset for constructing the feature dataset.


\subsection{Function-level Feature Extraction}
\label{subsec:feature_extaction}

Considering that SOTA code clone-based SCA tools for C/C++ repositories detect potential Type-2 code clones at the function level (i.e., CENTRIS~\cite{woo2021centris} and OSSFP~\cite{wu2023ossfp}). We follow this practice to construct a function-level feature dataset for SCA detection based on the selected 62,868 repositories.

Generally, we begin by cloning all selected repositories using the GitHub API~\cite{github-api}. For each repository, we identify functions with sufficient information and extract their features to construct the feature dataset. Specifically, we first collect all available GitHub versions and tags as snapshots of each repository. If a repository lacks version tags, we instead collect the main branch. Within each snapshot, we filter for C-related files. To minimize redundant processing, we leverage the observation that most files remain unchanged across version releases. Thus, for each snapshot, we retain only the C-related files that appear for the first time in the version history, effectively reducing duplication in feature computation.

Next, we extract functions from the C-related files in each snapshot of the selected repositories using ANTLR~\cite{antlr}. For each function, we remove comments and blank lines, and normalize function and variable names. Additionally, we filter out functions shorter than three lines, as they are considered insignificant~\cite{wu2023ossfp}. To ensure uniqueness, we generate an MD5 hash for each function and use these hash values to deduplicate redundant entries. As a result, we obtained 41,806,568 unique functions from 62,868 qualified repositories, forming the initial feature database.


\subsection{Greedy supplementary Strategy}
\label{subsec:supplementary}

Recognizing the potential limitations of initial metrics-based filtering, which may inadvertently exclude valuable repositories, we introduce a secondary step leveraging an anomaly detection methodology to recover such mis-filtered repositories. The rationale behind this approach is both straightforward and effective: a repository should be distinguishable from those that were excluded by demonstrating significantly stronger indicators. By modeling repositories within a vector space, we identify such repositories as outliers, allowing us to reintegrate them into our dataset.

Since the qualified repositories omitted in the initial filtering phase exhibit distinct characteristics, they can be regarded as global anomalies. Meanwhile, disqualified repositories tend to have a relatively concentrated distribution. Therefore, it is essential to select a detection algorithm that is sensitive to local outliers. To address this, we employ the Local Outlier Factor (LOF) algorithm, an unsupervised anomaly detection method specifically designed to identify local outliers within a dataset~\cite{lof}.

In particular, while the thresholds set in the Metric Filtering phase are already low (most are >0), we observe that some repositories display abnormally high values in one or more metrics. Compared to others, these repositories are more likely to attract user attention and be further reused, making them prime candidates for prioritization by the LOF algorithm.





To apply the LOF algorithm, all metrics are normalized and encoded into a six-dimensional vector as input. The algorithm begins by computing the k-distance for each repository vector and determining its k-nearest neighbors based on this distance. Next, it calculates the local reachability density (LRD) for each repository vector, which is derived from the reachability distance between the vector and its k-nearest neighbors. The LOF score is obtained by averaging the ratio of the LRD of a repository vector to the LRDs of its k-nearest neighbors. A threshold $\theta$ is defined to estimate the proportion of outliers. Based on our estimation, we set $k=10$ and $\theta=0.2$, ultimately identifying 469,883 outlier repositories.

\begin{table}[t]
    \centering
    \caption{Evaluation on Fitting the Feature Growth Curve}
    \label{tab:fit-model}
    \footnotesize
    \begin{threeparttable}
    \scalebox{0.95}{
        \begin{tabularx}{\columnwidth}{XXXXX}
            \toprule
            Model\tnote{*}   & RMSE     & MAE      & MAPE   & RMSRE    \\ \midrule
            Sigmoid          & 0.019068 & 0.012925 & 3.54\% & 0.126233 \\
            Exponential      & 0.033629 & 0.026435 & 6.22\% & 0.118601 \\
            Power Law        & 0.021906 & 0.017366 & 4.60\% & 0.140904 \\
            Hyperbolic       & 0.019855 & 0.014708 & 3.76\% & 0.079615 \\
            \multicolumn{1}{l}{Logarithmic}       &   \textbf{0.007187}   &  \textbf{0.005891}   &   \textbf{1.14\%} &   \textbf{0.020893}        \\
            \bottomrule
        \end{tabularx}
        }
    \end{threeparttable}
\end{table}

After identifying outlier repositories, we aimed to investigate an optimal stopping point for expansion. Instead of adding them all at once, we integrated them sequentially based on priority to monitor their impact. Outliers were batched by average rankings from basic metrics, and as each batch was added, we analyzed newly contributed unique functions. A change curve tracked feature diversity evolution, revealing a flattening trend, indicating diminishing returns from further additions.

To better understand and predict this trend, we explored fitting the curve using various mathematical functions, including the Sigmoid function, Exponential function, Power Law function, Hyperbolic function, and Logarithmic function, which have been widely applied in previous curve-fitting tasks~\cite{fitting-curve-1, fitting-curve-2, fitting-curve-3, fitting-curve-4}. To evaluate the fitting performance, we adopted root mean square error (RMSE), mean absolute error (MAE), mean absolute percentage error (MAPE), and root mean square relative error (RMSRE).

\begin{figure}[t]
    \centering
    \includegraphics[width=0.95\columnwidth]{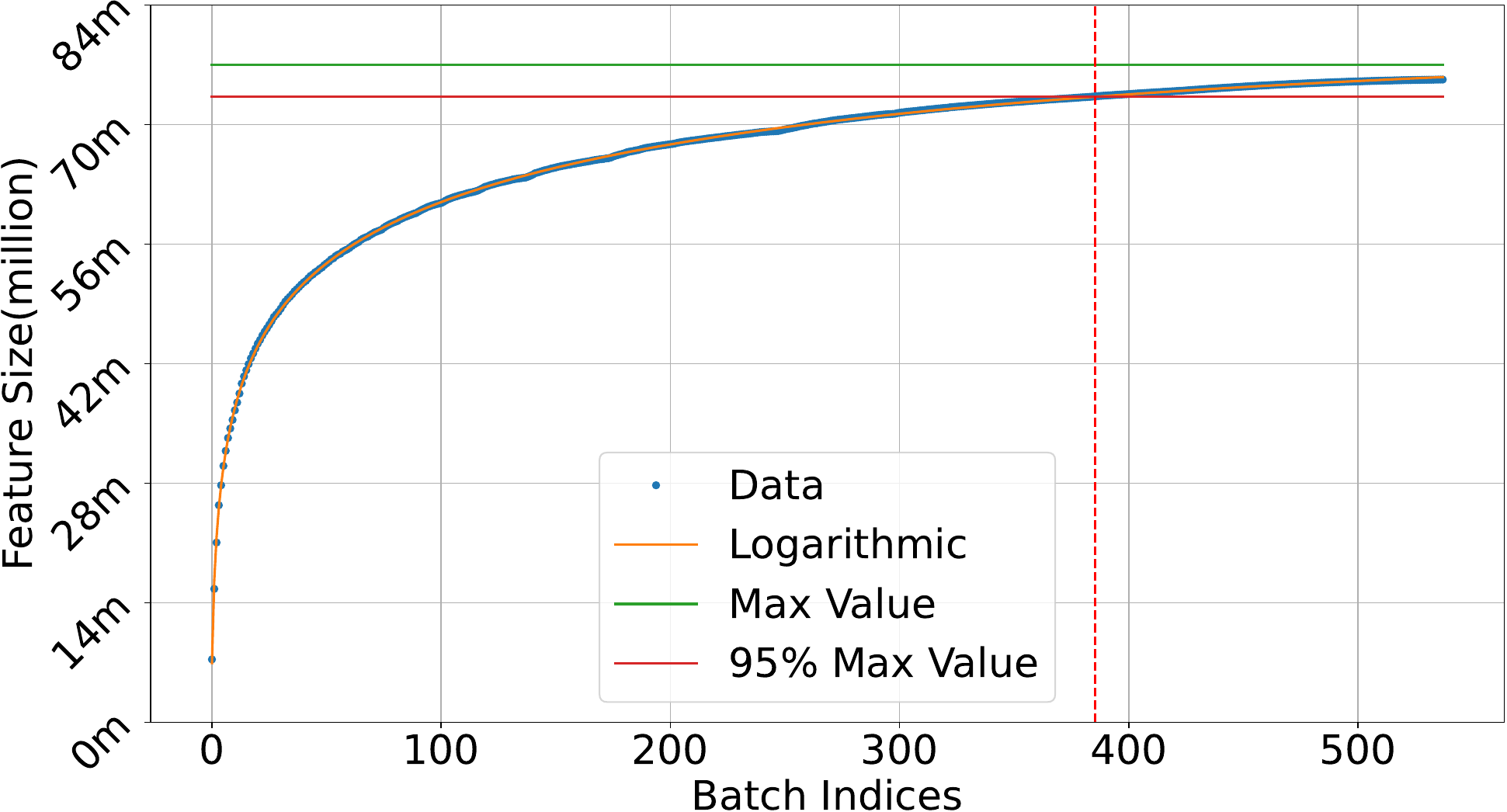}
    \caption{The amount of unique functions growing through per batch (each batch contains 1000 repositories).}
    \label{fig:feature_size_growing}
\end{figure}

The results are shown in \Cref{tab:fit-model}. The logarithmic function has the best performance in fitting the curve. We use the logarithmic function to estimate the growing curve. The fitted feature size estimation function is:
\begin{equation}
\begin{split}
\text{F}(x) = & \, 3.973 \times 10^{-1} \cdot \log(1.705 \times 10^{2} \cdot x + 1.235 \times 10^{2}) \\
                    & + 1.383 \times 10^{1} - 3.267 \times 10^{-4} \cdot x
\end{split}
\end{equation}
where $x$ presents the number of repositories we included, and $F(x)$ predicts the number of unique functions that are supposed to be retrieved from them. Given the vast landscape of 2.55 million C repositories hosted on GitHub, the fitted curve estimates approximately 77,650,766 unique functions across all these repositories, as illustrated by the green line in~\Cref{fig:feature_size_growing}. Based on this estimation, a subset comprising the top 385,000 repositories accounts for 95\% of these unique functions.

From a statistical analysis perspective, the concentration of functional features within a relatively small fraction of the total repository pool highlights the principle of diminishing returns when further expanding the dataset. Consequently, we established the cutoff for our supplementary data collection at this threshold of 385,000 repositories. The selected repositories within this scope are deemed sufficient to capture the diverse range of functional features inherent to C programming, providing a robust foundation for subsequent analysis and investigation.


\subsection{Low-value repositories filtering}

Although the 385,000 selected repositories can cover 95\% of all C language repositories on GitHub from the perspective of function diversity, the value attributes of the functions themselves are not considered. There are still repositories with low value for code reusing that are included in the scope. To this end, we refine our feature database by introducing several assumptions to exclude unsuitable repositories:

\begin{itemize}[leftmargin=*]
    \item \textbf{Function Quantity Extremes}: Repositories with either an excessively high or low number of functions, which often represent large-scale software or repositories of negligible relevance, are inappropriate for use as Third-Party Libraries (TPLs).
    \item \textbf{Function Repetition Rate}: Repositories with too high or too low function duplication rates will be excluded. The former means that the project cannot contribute enough unique functions and can be replaced by other repositories; the latter means that the project will basically not be copied.    
    \item \textbf{Complexity of Functions}: Repositories with functions that are either overly complex or simplistic are excluded. This criterion serves as an indirect measure of the repository's code quality.
\end{itemize}

We used the Maintainability Index~\cite{oman1992definition} to measure the complexity of the function. For each function, we extracted the following features to calculate MI.

\begin{itemize}[leftmargin=*]
    \item 
        \textbf{LOC (Lines of the code)}: The lines of function without blank lines, which is an important indicator of function complexity.
    \item 
        \textbf{Halstead Volume (HV)}~\cite{hv}: the measure of software size and complexity. It is based on counting a program's operators and operands, which is calculated by:
        \begin{equation}
            HV = (N_1 + N_2) \times \log_2(n_1 + n_2)
        \end{equation}
        $n_{1}$ and $n_{2}$ stand for the number of unique operators and unique operands in the program. $N_{1}$ and $N_{2}$ stand for the total number of operators and total number of operands in the program.
    \item 
        \textbf{Cyclomatic Complexity (CC)}~\cite{cc}: an indicator of complexity. It is based on the Control Flow Graph (CFG) of the function, which is calculated by:
        \begin{equation}
            CC = E - N + 2P
        \end{equation}
        $E$ stands for the number of edges in the CFG. $N$ stands for the number of nodes in the CFG. $P$ stands for the number of connected components of the function.
    \item
        \textbf{Maintainability Index (MI)}~\cite{mi}: the comprehensive measurement of the function complexity based on the comprehensive calculation of the above indicators.
        \begin{equation}
            MI = 171 - 5.2 \times \ln(\text{HV}) - 0.23 \times (\text{CC}) - 16.2 \times \ln(\text{LOC})
        \end{equation}
\end{itemize}

We addressed our assumptions by sorting repositories based on specific criteria and applying a filtering threshold $\theta$. For the first assumption, repositories were ordered by their total number of functions, with both the upper and lower extremes removed. The second assumption involved sorting by the proportion of duplicate functions and similarly excluding the top and bottom segments. For the third, we sorted based on the sum of MI of owned functions, again removing the extremes. The threshold $\theta$ controlled the extent of filtering, with its evaluation detailed in \Cref{subsec:effectiveness}.

\section{Experiment}
\subsection{Research Questions}

To evaluate the effectiveness of the constructed feature dataset, we try to investigate the following research questions:

\begin{itemize}[leftmargin=*]
    \item \textbf{RQ1: Effectiveness.} How can OSSScope improve the performance of state-of-the-art SCA tools?
    \item \textbf{RQ2: Ablation Study.} How does each step impact the final performance improvement?
    \item \textbf{RQ3: Characteristics.} What are the characteristics of repositories included/excluded by OSSScope?
    \item \textbf{RQ4: Representative.} How representative is the feature dataset generated by OSSScope compared to other data sources?
\end{itemize}


\subsection{Experiment Setup}

\noindent $\bullet$ \textbf{Ground Truth Dataset Preparation.} To address the challenge of identifying Third-Party Libraries (TPLs) without a standardized package manager or direct SBOM, we utilized a file-path mapping technique inspired by previous work. We filtered repositories based on paths containing keywords such as "/third\_party", "/3rdparty", and "/deps", identifying 3,386 TPLs across 1,663 repositories. Three software engineers and two researchers manually verified the metadata files for accuracy. 

Although we adhere to the ground truth labeling methodology established in previous works, we acknowledge its limitations, particularly the potential misidentification of TPLs due to their presence in different directory paths, which may result in false positives and false negatives. However, precise manual labeling would require substantial effort. To mitigate this threat, we conducted a further inspection of potential false positives, as detailed in ~\Cref{subsec:effectiveness}, and also addressed this issue in the threats section.

\noindent $\bullet$ \textbf{Evaluation Setting.} For evaluation, we re-implemented the OSSFP algorithm and compared its performance in version-level SCA on both our dataset and the original OSSFP dataset, which utilizes the same feature format (i.e., type-2 deduplicated function hashes). Since OSSFP is not open-sourced, we re-implemented the algorithm and reconstructed the dataset by strictly following the methodology outlined in the paper (collecting the repositories with over 100 stars). To ensure the correctness of our implementation and dataset construction, we contacted the authors for verification. 

Notably, we found that 86.9\% of the repositories in the OSSFP dataset (12,859 repositories with over 100 stars) are included in OSSScope's dataset, while the remaining repositories were filtered out due to the low-value repository filtering process.

\noindent $\bullet$ \textbf{Experiment Environment.} All experiments were conducted on an Ubuntu 20.04.6 LTS server with a 2.50GHz Intel Xeon Gold 6248 CPU and 188GB RAM. All SCA tools are set as default.


\subsection{RQ1:Effectiveness}
\label{subsec:effectiveness}

\begin{figure}[t]
    \centering
    \includegraphics[width=0.95\columnwidth]{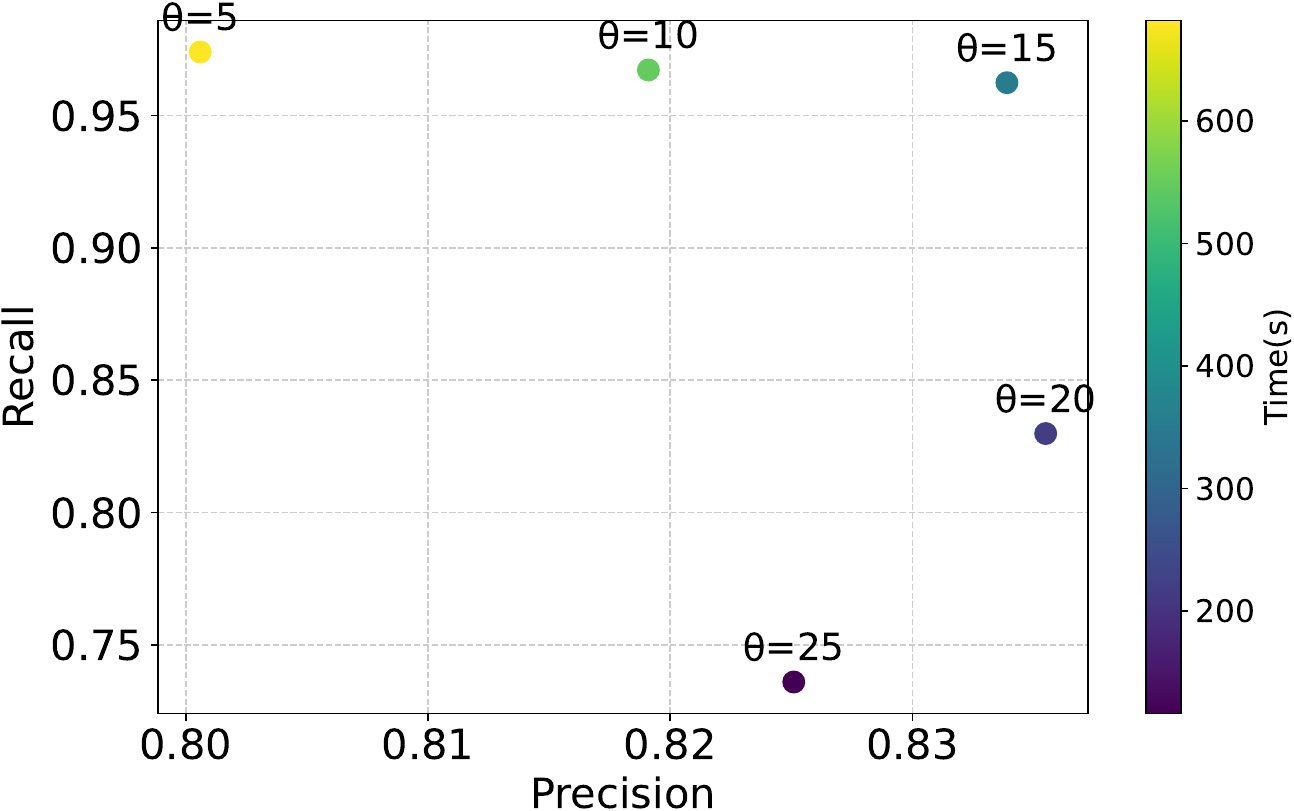}
    \caption{Precision, recall, and analysis time of different $\theta$}
    \label{fig:evaluation-theta}
\end{figure}

\noindent $\bullet$ \textbf{Threshold Selection.} Initially, we assessed various values of $\theta$ within the range of [5\%, 10\%, 15\%, 20\%, 25\%], corresponding to the exclusion of the top and bottom $\theta$ of repositories based on our predefined criteria. Utilizing the refined feature database, SCA experiments were conducted on the established ground truth. \Cref{fig:evaluation-theta} illustrates the outcomes in terms of precision, recall, and time cost associated with different $\theta$ values. As $\theta$ increased, the dimension of the selected repository pool and the corresponding feature database diminished, leading to reduced computational time. However, this reduction also entailed a decrease in recall, attributable to the omission of certain Third-Party Libraries (TPLs).

To strike a balance between precision and recall, we employed the $F_1$ score as a holistic metric for evaluation. After careful consideration, a $\theta$ value of 15 was selected, achieving the optimal $F_1$ score of 0.89, coupled with an acceptable average time cost of 0.21 seconds. This $\theta$ value represents an effective compromise, maximizing the accuracy of our SCA experiments while maintaining efficiency. The final feature database is constructed by 61,746,893 unique functions from 193,026 repositories.

\noindent $\bullet$ \textbf{SOTA SCA Tools Improvement}. We applied the OSSFP method to both OSSScope and its original feature database to evaluate the effectiveness across different feature databases using the ground truth dataset. The results are summarized in \Cref{tab:comparison_different_tools_on_ossscope}. Upon substituting the feature dataset, OSSScope achieved a precision of 83.39\% and a recall of 96.25\% on the ground truth dataset. This represents an improvement of 0.87\% in precision and a significant 22.65\% increase in recall compared to the performance metrics obtained with their original feature databases.

\begin{table}[t]
    \centering
    \caption{TPL Detection Results on Ground Truth}
    \label{tab:comparison_different_tools_on_ossscope}
    \footnotesize
    \scalebox{0.95}{
    \begin{tabularx}{\columnwidth}{lccccXXX}
        \toprule
        Feature    &  GT   &  TP   &   FP    &   FN   & Precision & Recall  & F1          \\ \midrule
        Original   &  3386 &  2492 &   528   &   894  &  82.52\%  & 73.60\% & 77.80\%     \\ 
        OSSScope  &  3386 &  3259 &   649   &   127  &  83.39\%  & 96.25\% & 89.36\%     \\ \bottomrule
    \end{tabularx}
    }
\end{table}

\noindent $\bullet$ \textbf{False Negative Analysis}. Compared to the original feature library, OSSScope has significantly reduced the number of false negatives. The primary reason is that the inclusion of third-party libraries (TPLs) is not limited to high-star repositories. While high-star repositories generally indicate greater recognition and visibility (e.g., they are prioritized in GitHub search rankings~\cite{github-search-api}), relying solely on this criterion overlooks repositories with lower evaluation metrics that are still widely used. Notably, 95.4\% (853 false negative cases) of the TPLs excluded by the original criteria belong to repositories with fewer than 100 stars.

However, false negative cases persist. A predominant issue stems from the use of timestamps as a criterion for determining clone relationships during function deduplication. Current state-of-the-art (SOTA) SCA detection methods infer import relationships based on submission time or commit time. However, the absence of a standardized versioning system across open-source platforms complicates this approach. Delays in marking submission times can introduce inconsistencies in the deduplication process. Consequently, even though OSSScope correctly incorporates the relevant TPLs, SCA algorithms that rely on timestamp-based criteria may inadvertently exclude these valid TPLs, leading to false negatives.

Another source of false negatives arises from ground truth labeling. Some TPLs residing in alternative directory paths may be overlooked by both the detection tools and the ground truth dataset. Nonetheless, restricting recall analysis to selected directory paths remains a reasonable approach for evaluation.

\noindent $\bullet$ \textbf{False Positive Analysis}. As the feature database expands, there is a notable increase in false positive cases. A significant contributing factor is the inclusion of certain "software families" within the database. These software families consist of variations of original software that have undergone secondary development. While they retain substantial code similarities, they differ in functional modules. Each family originates from a specific repository but evolves in different directions to serve distinct purposes. Consequently, repositories within the same family exhibit code duplication, which can be misinterpreted by the SCA algorithm, leading to false positives. 

Upon reviewing our false positive cases, we found that 15.71\% (102 out of 649) were attributed to software family similarities, including repositories derived from \textit{zlib} (15 cases), \textit{cJSON} (16 cases), \textit{jemalloc} (12 cases), \textit{linmath.h} (17 cases), among others. More details can be found on our website~\cite{public-data}. The concurrent inclusion of such closely related software in the feature library presents additional challenges. Due to the significant overlap in function features and potential inaccuracies in timestamp-based ordering—stemming from the platform's reliance on time tags that may not precisely reflect the actual sequence of code submissions.

To assess the threats associated with ground truth labeling, we conducted a deeper inspection of false negatives. Since some TPLs may reside in different directory paths, they should be included in the ground truth and detected as true positives. After manually reviewing these cases post-experimentation, we found that out of the 649 false positives, 37 were actually true positives that were omitted in the ground truth but correctly identified by our tools.

\begin{tcolorbox}[breakable, left=0pt, right=0pt, top=0pt, bottom=0pt, width=\linewidth, enlarge left by=0mm, boxsep=5pt, arc=0pt, outer arc=0pt,]
\textbf{Answer to RQ1:}
{Applying the same SCA algorithm, the function feature database provided by OSSScope outperforms the original method on the ground truth dataset, with precision and recall improving by 0.87\% and 22.65\%, respectively. Despite limitations imposed by the SCA algorithm's criteria, OSSScope enhances SCA performance through improved data quality.}
\end{tcolorbox}


\subsection{RQ2:Ablation Study}
\label{subsec:ablation}

To evaluate the effectiveness of the feature database generated by OSSScope, we conducted an ablation experiment to measure the impact of each step in the generation process. Different feature databases were constructed based on the intermediate results after executing each step, and we repeated the SCA experiments to assess the accuracy, efficiency, and time performance.

\begin{figure}[tb]
    \centering
    \includegraphics[width=0.95\columnwidth]{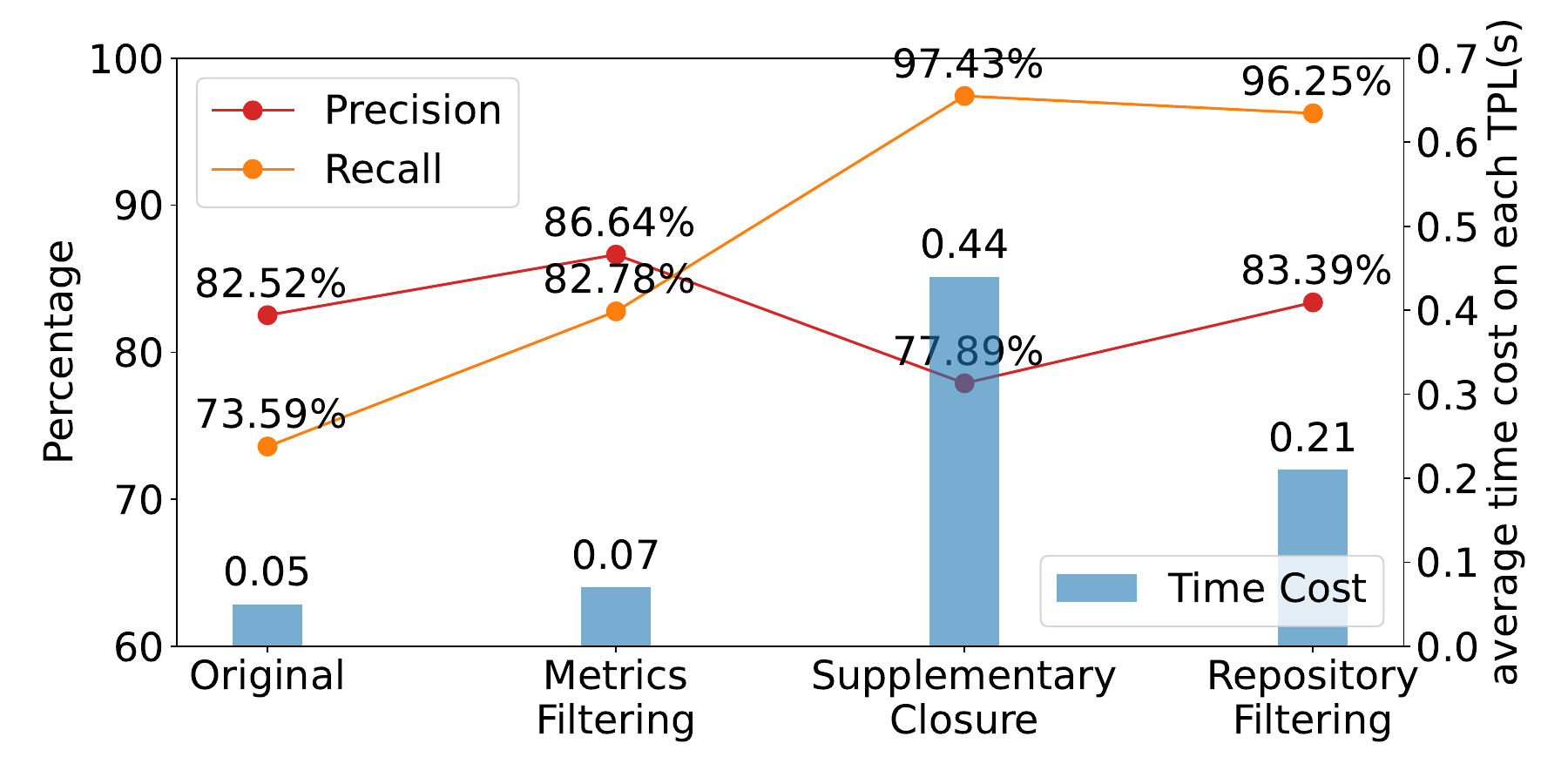}
    \caption{Accuracy and time cost among each step}
    \label{fig:ablation}
\end{figure}

\Cref{fig:ablation} illustrates the cumulative impact of each step on precision, recall, and average time cost, where the average time cost refers to the time on detecting each TPL component. Original stands for the feature from OSSFP. Initially, employing only metrics-filtered features improves precision and recall compared to the baseline, as qualified repositories prioritized at this stage include most TPLs. However, relying solely on this filtering can lead to false negatives by inadvertently excluding relevant repositories. To mitigate this, the supplementary strategy recovers mistakenly excluded repositories, reducing false negatives. However, as discussed in ~\Cref{subsec:effectiveness}, incorporating excessive features from "software families" complicates precise feature extraction, increasing false positives and prolonging processing times. The final repository filtering step refines OSSScope by eliminating redundant features, effectively reducing false positives without significantly impacting false negatives, as repositories filtered at this stage rarely contain TPLs. This step also reduces the feature database size, improving detection efficiency and lowering average processing time.

In terms of runtime performance, the original OSSFP dataset (12,859 repositories) requires 0.05 seconds per scan per TPL component. After applying metrics filtering (62K repositories), the time increases to 0.07 seconds. The supplementary closure step (385K repositories) significantly increases the time to 0.44 seconds, whereas the final repository filtering step (193K repositories) reduces the time to 0.21 seconds. This result underscores the critical role of the repository filtering step in balancing accuracy, feature database size, and computational efficiency.

\begin{tcolorbox}[breakable, left=0pt, right=0pt, top=0pt, bottom=0pt, width=\linewidth, enlarge left by=0mm, boxsep=5pt, arc=0pt, outer arc=0pt,]
\textbf{Answer to RQ2:} {The ablation study highlights the impact of each step on SCA results. Metrics filtering captures most TPL repositories, the supplementary closure recovers erroneously filtered ones, and repository filtering removes duplicates and unnecessary features. Together, these steps enhance feature quality while maintaining detection efficiency in OSSScope.}
\end{tcolorbox}


\subsection{RQ3:Characteristics}

{OSSScope has identified a representative subset of GitHub repositories to comprehensively capture the functional features. To further investigate the composition of the resulting feature database, we conducted an in-depth examination of the characteristics of both included and excluded repositories. By analyzing each metric individually, we quantified the numerical distribution of each metric. Additionally, we performed a semantic inspection of the excluded repositories to derive insights into their characteristics and potential reasons for exclusion.}



\noindent $\bullet$ \textbf{What kind of repositories are included by OSSScope.} For the size of the function, the selected repositories hold the number of functions ranged from 20 to 1,392. This range is considered appropriate for a repository to serve as a Third-Party Library (TPL). Additionally, these repositories have at least one star and exhibit activity spanning more than 10 days, aligning with the premise that actively maintained repositories are more likely to be reused.

Contrary to expectations, our analysis revealed many repositories with neither forks nor issues. Specifically, 87,524 repositories (45.3\%) lack forks, yet 7,479 of them have at least five stars. Similarly, 151,471 repositories (78.4\%) have no open issues, while 11,083 of them have at least five stars and five forks. This suggests a deviation from conventional user engagement practices in TPLs, where users may prefer copying source code directly rather than forking, and issue reporting is infrequent. Consequently, forks and issues are unreliable metrics for assessing a repository’s suitability as a TPL, highlighting the need for a more nuanced evaluation approach in feature database construction.

\noindent $\bullet$ \textbf{What kind of repositories are excluded by OSSScope.} {To gain insights into the characteristics of repositories excluded during the filtering step, we conducted a semantic analysis by statistically analyzing the word frequency in their descriptions. We tokenized the descriptions by splitting space and removing common stopwords, such as prepositions and particles. Since extreme values at both ends were excluded during repository filtering, we separately computed word frequencies based on the number of functions within each repository. Subsequently, we manually examined high-frequency words and categorized them into relevant topics.}

\begin{figure}[t]
    \centering
    \includegraphics[width=0.85\columnwidth]{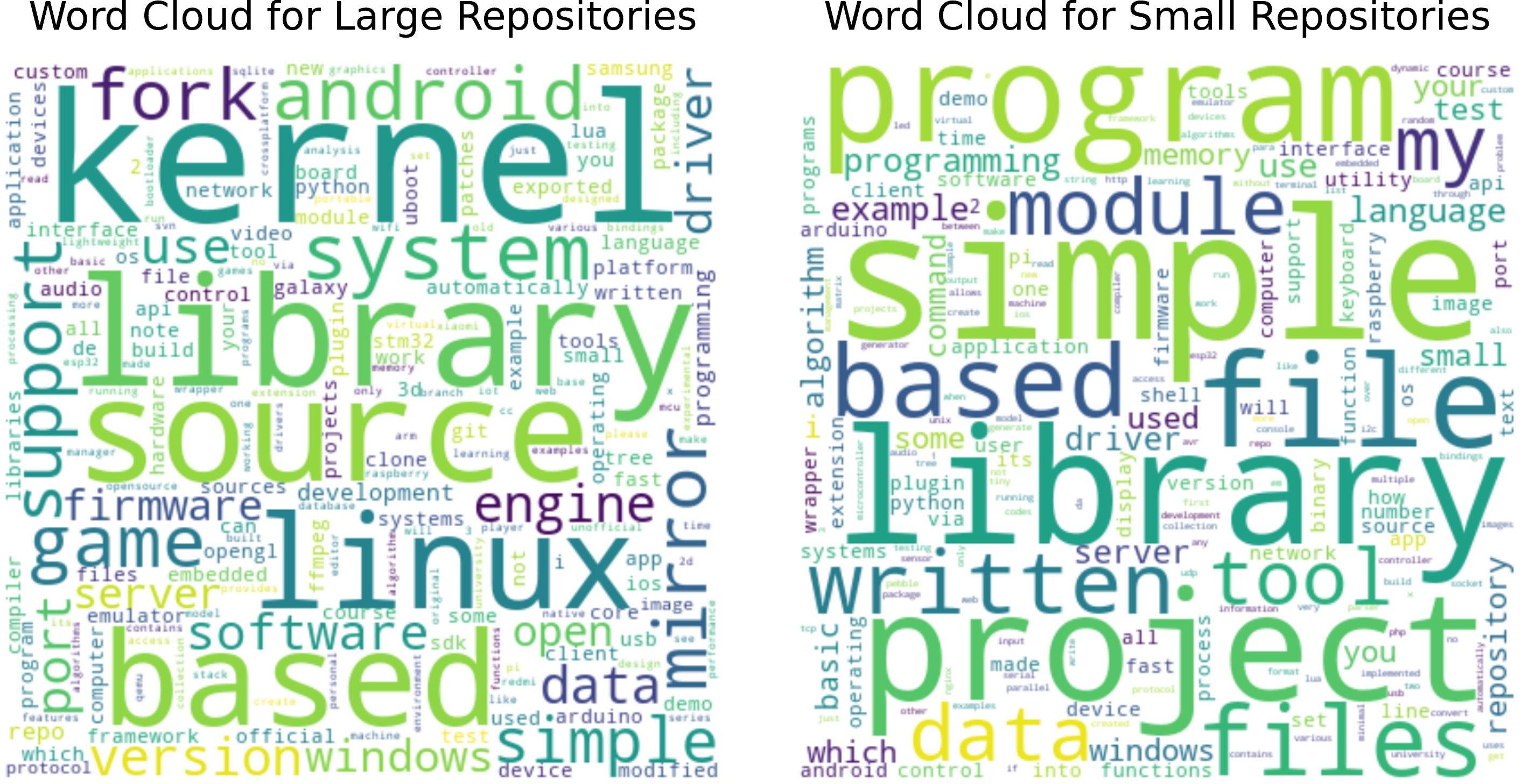}
    \caption{Word cloud of large/small repositories}
    \label{fig:word-freq}
\end{figure}

As shown in ~\Cref{fig:word-freq}, repositories with a large number of functions and high complexity frequently belong to categories such as operating systems, large-scale applications (e.g., games, Android), and decompiled code, aligning with common C language use cases. For instance, the Linux Kernel involves over 14,000 contributors. However, developers often bypass GitHub’s fork functionality, directly uploading modified versions as separate repositories, leading to redundancy. A notable example in game development is \textit{GTAV-Decompiled-Scripts}, which contains thousands of functions from decompiled game scripts. Despite its popularity, such repositories are unsuitable as TPLs. 

Conversely, repositories with a small number of functions and low complexity primarily fall into categories such as tutorials, examples, test cases, and software prototypes. While these repositories serve important educational and practical purposes and are widely recognized, they are commonly used as quick references or functional prototypes rather than true TPLs. Despite their utility and popularity, they do not align with TPL definitions or real-world usage scenarios and were thus excluded from the final feature library.

\begin{tcolorbox}[breakable, left=0pt, right=0pt, top=0pt, bottom=0pt, width=\linewidth, enlarge left by=0mm, boxsep=5pt, arc=0pt, outer arc=0pt,]
\textbf{Answer to RQ3:}
{The scope of repositories selected by OSSScope generally ranges from 20 to 1,392 functions, with the presence of forks and open issues not being a decisive factor. Additionally, we investigated the excluded repositories and identified six key topics, which provide valuable insights for improving the efficiency of repository filtering in future studies.}
\end{tcolorbox}


\subsection{RQ4:Representative}

To assess the representativeness of the feature dataset constructed by OSSScope, we conducted an evaluation using additional data sources. 
We understand that there are many repositories hosting C projects, we select 3 most commonly used repositories, i.e.,  
SourceForge~\cite{SourceForge}, GitLab~\cite{GitLab}, and Vcpkg.io~\cite{vcpkg}, which have also been constantly referenced by previous studies~\cite{tplite, wang2024preserving, wu2024packhunter}, to evaluate whether our constructed dataset covers features of C projects hosted on them.



For each of them, we randomly sampled 10\% of C repositories. Due to GitLab’s API limitations, which prevent language-based searches and restrict web access to 1,000 results, we capped the number of GitLab repositories in our evaluation at 1,000. Features were extracted from the latest available version of each repository. As shown in ~\Cref{tab:different-data-source}, OSSScope's dataset covers over 95\% of features from randomly sampled repositories across various sources. To assess representativeness, we analyzed repository origins. For Vcpkg.io, 98\% (251 cases) had corresponding or backup versions on GitHub, while for GitLab, 96\% (96 cases) had GitHub counterparts. For repositories not explicitly included in OSSScope, feature coverage analysis showed that most of their function features were already present in the dataset. This highlights OSSScope's diversity and its effectiveness in enhancing code clone-based SCA by improving clone function identification accuracy.

The primary reason that building a feature library based on GitHub is both comprehensive and effective is that GitHub is the most popular open-source platform, hosting a significantly larger volume of repositories than any other platform. Moreover, even for repositories that are not originally hosted on GitHub, developers or users frequently publish mirror versions or copies on the platform (e.g., Chromium is hosted by Google ~\cite{chromium-host}, it also has official mirror repository on GitHub~\cite{chromium}), further expanding its coverage. 

\begin{table}[t]
    \centering
    \caption{Diversity and Coverage on Different Data Source}
    \label{tab:different-data-source}
    \footnotesize
    \scalebox{0.95}{
    \begin{tabularx}{\columnwidth}{ccccc}
        \toprule
        Source & \#All C Repo & \#Samples & \#Total Features &   Coverage \\ \midrule
        SourceForge     & 27,198       & 2,720     & 715,363          & 95.17\%           \\
        GitLab          & 1,000        & 100       & 24,153           & 99.19\%           \\
        Vcpkg.io        & 2,562        & 256       & 59,904           & 97.42\%           \\
        \bottomrule
    \end{tabularx}}
\end{table}

We acknowledge that relying solely on GitHub for dataset construction may result in the omission of certain repositories or features. However, collecting this small fraction of additional data would require substantial effort, as there is no unified or standardized method for accessing repositories hosted on decentralized sources, such as personal developer homepages~\cite{gtkmm}. OSSScope strikes a balance between collection overhead and dataset quality, ensuring a representative and diverse feature set while maintaining feasibility in data acquisition.

\begin{tcolorbox}[breakable, left=0pt, right=0pt, top=0pt, bottom=0pt, width=\linewidth, enlarge left by=0mm, boxsep=5pt, arc=0pt, outer arc=0pt,]
\textbf{Answer to RQ4:}
{The feature dataset from OSSScope covers over 95\% of function features from other sources, ensuring diversity and representativeness. While some repositories remain uncollected, their limited unique contributions and high acquisition costs make inclusion impractical. OSSScope efficiently leverages GitHub, balancing data collection efficiency and dataset quality.}
\end{tcolorbox}

\section{Migration to Other Language Ecosystem}
\label{sec:migration}

Our approach on SCA feature dataset construction has been proved to be effective by previous experiments. However, Given that applying OSSScope requires substantial time, storage, and computational resources (e.g., downloading source code and calculating incremental features), it is still too heavy to construct such a dataset (i.e., it took us 4 months for the construction on an 80-core and 188GB-RAM server). To this end, in this section, we aim to steam the insights from our results and introduce a more lightweighted way for SCA feature database construction in other ecosystems. 

Specifically, based on the metrics outlined in \Cref{tab:criteria}, we propose a bi-classification machine learning model,RClassifier, to classify whether a given repository, with certain combination of these metrics, should be selected to construct SCA feature databases, trained by the inclusion and exclusion of each C repository by our feature database for C.
RClassifier enables repositories to be filtered directly based on meta-level metrics, significantly reducing the time and computational overhead associated with feature analysis. To evaluate its effectiveness, we applied RClassifier to another widely used ecosystem, Golang, which is approximately half the size of the C language ecosystem.

\subsection{The Lightweight Method: RClassifier}

For this binary classification task (i.e., determining whether to include a repository), we selected lightweight and well-established classical models commonly used in previous studies~\cite{kumari2017machine,jeatrakul2009comparing,bounsiar2008general}. These models include Logistic Regression (LR), Support Vector Machines (SVM), Decision Trees (DT), Random Forest (RF), Gradient Boosted Machines (GBM), Naive Bayes (NB), K-Nearest Neighbors (KNN), and Linear Discriminant Analysis (LDA). We utilized the results from \Cref{sec:exploratory} as the training dataset, comprising 193,026 positive samples (repositories selected by OSSScope) and 2,356,855 negative samples (the remaining C repositories). The models were trained using 10-fold cross-validation~\cite{k-fold}. Model performance was evaluated by precision, recall, and $F_1$ scores.

\begin{table}[t]
    \centering
    \caption{Evaluations of different models}
    \label{tab:model}
    \footnotesize
    \scalebox{0.95}{
    \begin{tabularx}{\columnwidth}{lXXX}
        \toprule
        Model    &  Precision  &  Recall    &  $F_1$ Score     \\ \midrule
        Logistic Regression       &   81.41\%   &  63.10\%   &   71.10\%        \\   
        Support Vector Machines      &   76.87\%   &  76.11\%   &   76.49\%        \\  
        Decision Trees       &   94.65\%   &  95.30\%   &   94.97\%        \\
        \multicolumn{1}{l}{\textbf{Random Forest}}     &   \textbf{94.65\%}   &  \textbf{96.52\%}   &   \textbf{95.58\%}        \\
        Gradient Boosted Machines      &   88.13\%   &  85.47\%   &   86.78\%        \\  
        Naive Bayes       &   92.90\%   &  30.33\%   &   45.72\%        \\ 
        K-Nearest Neighbors      &   83.63\%   &  86.95\%   &   85.26\%        \\  
        Linear Discriminant Analysis      &   82.87\%   &  57.58\%   &   67.95\%        \\
        \bottomrule
    \end{tabularx}
    }
\end{table}

The results for each model are presented in \Cref{tab:model}, with Random Forest emerging as the best-performing model. We named this model RClassifier and applied it to other language ecosystems. While we acknowledge that other models might offer performance improvements, the selected model has demonstrated sufficient effectiveness in generalizing repository selection for SCA feature dataset construction.

\subsection{Metrics Importance Analysis}

Given the Random Forest algorithm's capability to quantify the importance of features, we computed the importance weights of each feature in relation to the prediction outcomes. By observing how model performance varies with random perturbations of feature values, we identified the most influential factors contributing to the prediction results.

As shown in \Cref{tab:feature-importance}, action period, number of commits, and number of issues were the most critical features, together accounting for a significant portion of the overall importance. These factors reflect the activity level and engagement within a repository. Specifically, projects with longer activity periods and frequent commits demonstrate ongoing maintenance and development, which provides a rich and diverse set of code changes. This diversity enhances the feature set's representativeness for clone detection tasks. Additionally, a high number of issues indicates strong community interaction and suggests that the project addresses various practical problems, further enriching the potential for clone detection. In contrast, the number of contributors, number of stars, and number of forks were less influential. While these features reflect a project’s popularity and collaborative nature, they do not directly correlate with the volume or complexity of code changes, which are more pertinent to building an effective feature database for clone detection. For instance, stars are often an indicator of external recognition rather than code evolution (e.g., well-written tutorials and demos), and contributors may vary in the significance of their contributions.

\begin{table}[t]
    \centering
    \caption{Importance of each feature on RClassifier}
    \label{tab:feature-importance}
    \footnotesize
    \scalebox{0.95}{
    \begin{tabularx}{\columnwidth}{XXXX}
        \toprule
        Feature        &  Importance   &  Feature        &  Importance   \\ \midrule
        \#Stars        &   11.28\%     &  \#Contributors &   11.45\%    \\   
        \#Forks        &   10.56\%     &  Action Period  &   24.58\%    \\
        \#Issues       &   20.36\%     &  \#Commits      &   21.77\%     \\ 
        \bottomrule
    \end{tabularx}
    }
\end{table}

\subsection{Migration Performance of OSSScope and RClassifier on Go Ecosystem}

To evaluate the performance, we adopted OSSScope and RClassifier on the Go ecosystem, respectively. For OSSScope, we followed the procedures introduced in \Cref{sec:exploratory} and kept all the parameters the same. The constructed feature dataset contained 47,871,633 features from 93,852 repositories. Then, using RClassifier, we generated a repository list with 87,271 repositories directly based on the pre-collected meta information. Based on these, we constructed a function feature database with 46,739,826 function features. 
As for the ground truth, we manually marked the ground truth in the same way and conducted an SCA experiment to compare the performance of the two feature datasets.

\begin{table}[t]
    \centering
    \caption{TPL Detection Results on Go Language Ecosystem}
    \label{tab:go_sca_result}
    \footnotesize
    \scalebox{0.95}{
    \begin{tabularx}{\columnwidth}{XXXX}
        \toprule
        Feature     &   Precision &  Recall   & $F_1$ Score       \\ \midrule
        OSSScope   &   86.69\%   &  89.67\%  & 88.15\%           \\ 
        RClassifier  &   84.61\%   &  92.54\%  & 88.40\%           \\ \bottomrule
    \end{tabularx}
    }
\end{table}

As shown in \Cref{tab:go_sca_result}, the results indicate that RClassifier improves precision by 2.08\%, slightly reduces recall by 2.87\%, and maintains the same $F_1$ score as the full calculation process. In terms of efficiency, RClassifier determines the repository scope within seconds using only meta-information, whereas the full process requires several weeks. Despite a slight reduction in recall, RClassifier substantially enhances computational efficiency, making it a practical alternative when resources or time are limited. While the full process provides a more exhaustive analysis, the model-based approach offers considerable time savings, making it particularly useful for large-scale analyses where downloading and processing source code is impractical. This trade-off enables users to adjust between accuracy and efficiency based on specific task requirements.

A deeper analysis of the performance gap reveals that RClassifier performs well for repositories with strong metric signals but struggles with borderline cases. OSSScope addresses this issue by integrating incremental feature analysis and rule-based filtering, allowing it to distinguish ambiguous repositories with greater accuracy. In contrast, RClassifier lacks these refined semantic insights, limiting its effectiveness in classifying borderline cases. Nevertheless, RClassifier remains highly effective in filtering out non-relevant repositories based on readily available metadata, significantly reducing computational overhead compared to OSSScope.

\section{{Implications}}
\label{sec:discussion}

\noindent $\bullet$ \textbf{Generalization}. The proposed methods, OSSScope and RClassifier, have demonstrated their effectiveness in constructing a comprehensive CCD feature dataset and efficiently adapting to other programming languages. However, there are still two key directions for enhancing their generalization ability. 

Different programming languages have distinct grammatical structures and development conventions. \textbf{1)} To enhance adaptability, language-specific rules should be established, considering variations in function feature granularity, syntax parsing, selected metrics, and thresholds. Refining these parameters would enable OSSScope and RClassifier to better adapt to diverse language ecosystems, ensuring accurate feature extraction. \textbf{2)} Additionally, the current feature dataset treats all features and repositories equally for general CCD tasks. In domain-specific applications, prioritizing relevant features or repositories could further optimize performance. Implementing a ranking mechanism or secondary filtering could reduce dataset size while preserving critical features, enabling a more targeted and efficient CCD process.
By addressing these two directions, OSSScope and RClassifier can be further refined to enhance their applicability across multiple programming languages and adapt to specific CCD use cases, ultimately improving both precision and computational efficiency.

\noindent $\bullet$ \textbf{Future Directions for Enhancing SCA}. In SCA experiments, misidentifications by the clone detection algorithm result in false positives and false negatives. Despite incorporating original TPLs into the feature dataset, inaccurate versioning due to incorrect timestamps can lead to erroneous clone relationships. In cases of secondary development, inherited repositories imported as TPLs may cause duplicate code to be incorrectly attributed to the original repository, further increasing false positives.

To mitigate these issues, \textbf{1)} a robust version tagging mechanism is required to ensure that versions accurately reflect key development milestones through comprehensive code volume and semantic analysis. Refining versioning strategies can minimize clone misattributions and improve TPL detection accuracy. \textbf{2)} Additionally, integrating semantic features, such as code context and documentation, can further enhance clone detection precision. \textbf{3)} Furthermore, the constructed feature dataset is significantly larger than existing datasets, relying solely on concise hash features. This scale poses challenges to both matching accuracy and computational efficiency. Enhancing matching strategies, refining indexing mechanisms, and developing scalable detection techniques are essential to maintaining the effectiveness of SCA in real software ecosystems.

\noindent $\bullet$ \textbf{Implications of the constructed dataset}.  
Apart from serving as comprehensive feature databases for SCA tools, the well-constructed dataset can also be valuable and fundamental to foster further research.
\textbf{1)} Our dataset naturally records the functions that are reused in different repositories, which could be a valuable analysis dataset for ecosystem-wide analyses, such as vulnerability propagation, and license noncompliance by code clones.
\textbf{2)} Repositories with high code-sharing ratios also tend to be derived from the same software ancestors. Based on our dataset, we can easily identify the potential migration relations of repositories, and provide a complete dataset for analysis of software evolution across repositories, as well as the migration analysis of specific risks and manipulations, such as vulnerabilities, patches, PoCs, code refactoring, etc. These data could be valuable for many specific SE tasks (i.e., similar to what the Software Heritage Project intended to~\cite{heritage}).
\textbf{3)} Function clusters that often appear in the same repositories could be modules that are widely used third-party components. Our dataset is also capable of identifying third-party libraries that should be indexed and protected by the open-source community. 
\textbf{4)} Our dataset can also be a well-deduplicated dataset for LLM training with maximum diversity, which could be incorporated to improve the training efficiency of CodeLLMs.

\section{{Limitations and Threats to Validity}}
\label{sec:threats}

We summarized the limitations and threats of the proposed methods, which can be improved in the future. \textbf{1)} The repository filtering metrics may introduce bias, as they do not directly reflect the likelihood of reuse. To mitigate this threat, we conducted a survey and selected metrics based on the summarization of common practices. \textbf{2)} The adoption of the LOF algorithm for filtering disqualified repositories was based on the observed distribution of the selected metrics, which may introduce bias by omitting certain repositories. Regarding the stopping threshold (i.e., 95\%), it was determined based on a cost-benefit trade-off. The remaining 5\% of features may contain valuable items. Although our evaluation demonstrated the effectiveness and representativeness of the constructed dataset, the absence of these features could still impact the detection accuracy of SCA. \textbf{3)} We formulated three hypotheses to exclude low-value repositories. As these aspects have not been extensively studied, we derived them based on empirical observations. While we analyzed patterns and insights from the excluded repositories, further validation is required to ensure the completeness and accuracy of these filtering rules. \textbf{4)} Given the existence of multiple open-source platforms, GitHub alone cannot fully represent all C repositories. Our experiments showed that the constructed dataset covered 95\% of the features compared to other sources. However, the remaining 5\% features could still affect the accuracy of SCA. \textbf{5)} Our ground truth labeling followed methodologies from previous studies. However, considering only the target directory path may introduce false positives and false negatives, as TPL reuse is often more flexible in practice. Although we manually inspected SCA results and analyzed potential biases in the ground truth, further improvements can be achieved by introducing a more precise labeling methodology.

\section{Conclusion}
\label{sec:conclusion}
In this paper, we proposed a systematical approach named OSSScope to construct a function feature dataset for SCA. Utilizing the generated features has improved the accuracy of the SOTA SCA tool. We also provide a lightweight model, named RClassifier, to migrate to other language ecosystems. Compared with the traditional method that only relies on simple metrics, OSSScope offers a more efficient method of constructing high-quality feature datasets.

\section*{Data Availability}
The constructed function feature database and the related scripts are published in \cite{public-data}.

\bibliographystyle{IEEEtran}
\bibliography{myref}

\newpage

\end{document}